\documentclass[11pt]{article}

\usepackage{graphicx}
\graphicspath{{images/}}
\usepackage{xspace}
\usepackage{subfigure}
\usepackage{booktabs}
\usepackage{comment}
\usepackage[left=2.5cm,right=2.5cm,top=2.5cm,bottom=2.5cm]{geometry}

\usepackage[sort&compress,numbers]{natbib}
\bibpunct{[}{]}{,}{n}{,}{,} 
\usepackage{fancyhdr}
\fancypagestyle{plain}{ %
\fancyhf{} 
\lhead{\textit{Form-free size distributions from SAXS/TEM}}
\rhead{J. M. Rosalie \& B. R. Pauw }
\lfoot{\textit{Author accepted manuscript}}
\setlength{\footskip}{45pt}
\rfoot{\thepage}
}

    \makeatletter
    \let\@fnsymbol\@arabic
    \makeatother

\newcommand{\betap}{\ensuremath{{\beta_1^\prime}}\xspace}

\title{Form-free size distributions from complementary stereological TEM/SAXS on precipitates in a Mg-Zn alloy}
\author{
Julian M. Rosalie
\thanks{Structural Materials Unit, National Institute for Materials Science (NIMS),
1-2-1 Sengen, 305-0047, Tsukuba, Japan. $\mathrm{\langle rosalie.julianmark@nims.go.jp \rangle}$} 
\and  
Brian R. Pauw\thanks{International Center for Young Scientists (ICYS)
National Institute for Materials Science (NIMS),
1-2-1 Sengen, 305-0047, Tsukuba, Japan. $\mathrm{\langle brian@stack.nl \rangle}$} 
}
\date{}

\usepackage[plainpages=false,
pdftitle={Form-free size distributions from complementary stereological TEM/SAXS on precipitates in a Mg-Zn alloy},
pdfauthor={J. M. Rosalie,  B. R. Pauw},
pdfsubject={Form-free size distributions from complementary stereological TEM/SAXS on precipitates in a Mg-Zn alloy},
pdfkeywords={Stereology,Small angle X-ray scattering (SAXS),Transmission electron microscopy (TEM),Precipitation,Magnesium-zinc alloys},
colorlinks,
debug,
linkcolor=red,
citecolor=red, 
filecolor=blue,
breaklinks=true]{hyperref}
\usepackage{doi} 

\begin{document}

\maketitle



\begin{abstract}
This work describes a multidisciplinary research methodology for quantifying the size distribution of nanoscale precipitates in polycrystalline alloys. Complementary transmission electron microscopy (TEM) and small-angle x-ray scattering (SAXS) are employed in a study of precipitate growth in an isothermally aged Mg-Zn alloy. 
TEM is used to identify the precipitate phases as rod-shaped \betap particles and to determine their radii and aspect ratio. Subsequently, SAXS data obtained from bulk quantities of the alloy is interpreted via a novel Monte-Carlo method to obtain accurate, form-free size distributions. Good agreement was obtained between particle radii distributions measured by both methods, exemplifying the applicability of this complementary methodology to study precipitation in textured alloys containing particles anisotropic with well-defined orientation-relationships to the matrix. 
\end{abstract}
\enlargethispage{1cm}

\section{Introduction}

Nanometer-scale precipitates are known to drastically affect mechanical properties of alloys. Accurate characterisation of the size distributions and volume fractions involved through accurate three-dimensional measurements are therefore crucial for establishing structure-property relationships essential for further material development. Such structural information can improve understanding not only of the precipitation kinetics, but also its relation to the strength \cite{Nie1999,ZhuAluminium2002,NieMg2003,RosalieMgZn2012,RosalieMgZnY2013}, fracture behaviour \cite{LiuZhang2004} and interface structure \cite{ArdellInterface2005}.

Transmission electron microscopy (TEM) is  traditionally used for analysis of such precipitates in alloys \cite{LiuReview1993} and has frequently been applied in studies of precipitates in  aluminium \cite{Karlik2004,DeschampsLae2007,Fang2012} and magnesium alloys \cite{RosalieMgZn2012,RosalieMgZnY2013} in addition to steels \cite{Allen-1993} and Ni-based superalloys \cite{Risbet2008}. 
TEM stereology involves  acquiring projected images of particles  ranging from nanometer to micron size in thin foils and using analytical  relationships \cite{underwood:1970,dehoff:1982,kelly:1982} to deduce particle size, volume fraction and number density etc. 
Unfortunately, TEM is restricted to small volumes and hence low precipitate populations, making it poorly suited for statistical analysis \cite{Geortz2009}. 
This is primarily due to the limited foil thickness, which is constrained by the need to maintain electron transparency. 
The need to minimise overlap in projection can also be problematic for dense particle distributions. 
In addition, TEM has remained resistant to automation and remains a labour-intensive and time-consuming task. 

Nanoscale particles are also amenable to characterisation by small-angle scattering (SAS), in particular small-angle neutron- and X-ray scattering (SANS and SAXS, respectively). SAS can characterise nanostructural features with dimensions identical to TEM. The technique has been reviewed by (amongst others) Fratzl \cite{Fratzl2003} and Kostorz \cite{KostorzSaxs2001} and its application to alloys is discussed by Triolo et al. \cite{Triolo1989b}.
Some practical benefits of SAS are that a minimum of sample preparation is needed \cite{Geortz2009}, and that the greater beam diameter and higher penetration depth of neutrons and X-rays probe a vastly larger volume (typically 1\,cm$^{3}$ for neutrons and between 10\,$\mu$m$^{3}$--1\,mm$^3$ for X-rays), which is more representative of the bulk material. The main challenge in the application of SAS lies in the complex data interpretation required for the extraction of physically meaningful parameters. This is partially due to the information loss inherent in the collection of SAS data but can also be ascribed to the complexities of analysis for anything but the simplest of scattering behaviour \cite{Pauw-2013a}. It is therefore common to resort to matching the scattering pattern against one of several highly idealised formalisations, with strict assumptions about the nature of the distribution \cite{Pedersen-1997}. As this work focuses on X-ray scattering, we shall henceforth focus on discussing SAXS.

Many previous metallurgical SAXS studies are therefore focused on cases where the precipitates are essentially globular in shape, spheres being one of the simplest scattering geometries to analyses. These include studies on Guinier-Preston zones (GPZ) in Al-Zn-Mg \cite{Bardhan1968,Deschamps:2001,Deschamps1999} and Al-Ag alloys \cite{dubey:1991,erni:2003,malik:1997}, spherical $\delta^\prime$ precipitates in Al-Li alloys \cite{PletcherAlLi2012} and carbides in carbon steels \cite{Allen-1993}. Numerous studies employing a combination of SAXS and TEM have also concentrated on essentially isotropic scatterers  \cite{Geortz2009,DeKruif1988,Pedersen1994,Mori2006,DeMoor1999,Pedersen1996,Schnepp-2013}, or low aspect ratio particles 
(e.g. cuboidal precipitates in Ni alloys \cite{Sequeira-1997}). The scattered intensity from these scatterers can be considered independent of the crystallographic texture and crystallographic orientation relationship between the matrix and precipitate, greatly simplifying the data analysis. High-aspect ratio precipitates, however, are of increased interest in the metallurgical field as they can affect the properties much more strongly than globular precipitates can for a given volume fraction.

Good studies of high-aspect ratio particles do exist (See, for example \cite{Fratzl2003,Sequeira-1997}) but these are commonly restricted to highly idealised distributions, and it is not uncommon to describe even polydisperse systems using only a single size parameter (e.g. a ``Porod'' radius) \cite{Rashkova2008}.
Moreover, the crystallographic texture and precipitate-matrix orientation relationship will strongly influence scattering from anisotropic particles in polycrystalline samples, but these factors are frequently overlooked. 

A practical data interpretation methodology based on a Monte-Carlo optimisation process has recently been developed, offering a conceptually straightforward determination of \emph{form-free} particle size distributions that can be easily extended to support a variety of particle shapes \cite{Pauw-2013, Schnepp-2013}. 
Given supporting information on the precipitate morphology and orientation (which TEM can provide), this method should be able to extract a unique solution for the size distribution of aligned, anisotropic particles. 

This work sets out to apply TEM and SAXS using the Monte Carlo analysis method for size distribution retrieval in a model system containing non-globular, high-aspect ratio scatterers embedded  in a polycrystalline matrix. Key characteristics required in such a model system are: 1) a known crystallographic texture, 2) a clearly defined precipitate morphology, 3) a known crystallographic orientation relationship between the precipitate and matrix, 4) a precipitate dimension amenable to both techniques and 5) industrial relevance.

Magnesium-zinc (Mg-Zn) alloys provide an ideal model system.  As is common with magnesium alloys, Mg-Zn alloys develop strong texture when extruded. These alloys derive high strength from a fine dispersion of high aspect-ratio rod-like precipitates (termed \betap) aligned parallel to the hexagonal axis of the magnesium matrix \cite{Sturkey1959}. These precipitates are the principal contributor to the strength of the alloy and also control the ductility \cite{RosalieMgZnY2013}. Mg-Zn alloys also form the basis of the industrial ``ZK'' series of alloys  and establishing an accurate correlation between the physical properties and the precipitate morphology, size distribution and volume fraction is therefore of practical importance.

In this work, a magnesium-zinc alloy is extruded to develop texture and then isothermally aged in order to produce an alloy containing aligned, high-aspect ratio particles. The \betap precipitate size distribution and volume fraction are measured using TEM and small-angle X-ray scattering. TEM provides direct information (size, orientation, morphology etc.) on a limited number of particles, providing the necessary constraints needed to obtain a unique, physically realistic solution from SAXS. Monte Carlo SAXS analysis subsequently provides bulk values for volume fraction and form-free size distributions. This multidisciplinary approach is supplemented by electron backscattered diffraction (EBSD) to characterise the texture and by microindentation hardness testing to monitor the precipitation strengthening response. 

\section{Experimental}

A binary magnesium-zinc alloy was prepared from high-purity elements by direct chill casting.
A  composition of Mg-3.4at.\% Zn was measured  by atomic emission spectroscopy.  
The billet was homogenised at 300$^\circ$C for 24\,h and extruded at 300$^\circ$C with an extrusion ratio of 12:1 to form a 12\,mm diameter rod.
Samples were cut from this rod using a slow-speed diamond saw, encapsulated in argon, solution-treated at 300$^\circ$C for 1\,h and quenched into water at  ambient temperature.

The microstructure and texture were examined in the solution-treated condition using light microscopy and electron back scattered diffraction (EBSD). 
Reflected light micrographs were obtained from ground and polished samples both in cross-section (i.e. normal to the extrusion direction (XD)) and longitudinal section (parallel to the XD). 
An acetic-picral etchant was used to highlight the grain boundaries.

The texture of a solution-treated, cross-sectional sample was measured via  electron back scattered diffraction (EBSD) using a field-emission scanning electron microscope (FE-SEM). 
The sample was prepared by grinding  (to 4000 grit) and polishing with 6$\mu$m and 1$\mu$m diamond paste. 
Final polishing made use of colloidal silica.
The data from this microscope was analysed using OIM Analysis software, version 6.1. 

The ageing response was examined by microindentation hardness testing.
Cross-sectional samples were isothermally aged at 150$^\circ$C in a silicon oil bath and quenched in water at ambient temperature. 
Each sample was then embedded in a low-temperature resin, ground and polished. 
The Vickers micro indentation hardness was measured using a Matsuzawa AMT-X7FS automated tester.
Each test included a total of 16 measurements per sample using a load of 100\,g
(ST) or 200\,g (all other samples). 
Previous studies reported maximum hardness for Mg-Zn alloys after 48\,h ageing at this temperature \cite{RosalieMgZn2012}, hereafter referred to as ``peak age''.
Ageing times in used this work range from 0\,h (not aged, referred to as ``solution treated'' or ST) to 406\,h (significantly overaged). 

\subsection{SAXS measurements}\label{sc:saxsmeas}

The SAXS measurements are performed on a laboratory instrument based around a rotating anode generator with a molybdenum target. Mo\,$k_{\alpha,1}$  radiation was selected and focused using an Osmic confocal mirror. The incident beam was further collimated using three pinholes of 0.3, 0.2 and 0.45\,mm in diameter, with the sample positioned immediately after the third pinhole. The distances from the optics to pinhole \#1, \#2 and \#3 are 0.115, 0.582 and 0.86\,m, respectively. Assuming that the second pinhole is the beam-defining pinhole, the transverse coherence length (and therefore the maximum size of precipitates contributing to the scattering signal) is on the order of 200\,nm \cite{Veen-2004}. Nominal beam diameter at the sample position is estimated to be approximately 200\,$\mu$m. The flightpath is a continuous vacuum from optics to detector.

A Dectris Pilatus 100k detector is used in combination with a translation stage to detect scattered radiation. Three images are collected for each measurement and stitched together to form a single, almost square scattering image consisting of 487 by 497 pixels, each measuring 172\,$\mu$m by 172\,$\mu$m. The detector is placed 1.36\,m from the sample, its distance verified by ruler and measurement of a silver behenate standard, allowing for an angular coverage of $1\times10^8$ to $3.7\times10^9$ reciprocal meter ($0.1$ to $3.7$\,nm$^{-1}$, respectively) in scattering vector $q$. $q$ is defined as $q=4\pi/\lambda \sin(\theta)$, where the radiation wavelength is denoted as $\lambda$, and $2\theta$ is the scattering angle.

SAXS measurements are performed on cross-sectional samples manually ground to a thickness of $\sim$0.6\,mm. Final grinding was performed using 4000 grit paper to suppress scattering from surface features, and the samples were stored and measured in vacuum to prevent surface oxidation. The samples were measured with the X-ray beam parallel to the extrusion axis. The scattering patterns showed an overall isotropic scattering pattern, indicating that the samples were measured through the centre section of the slice (c.f. Paragraph \ref{sc:texture}). 

The data correction methodology is detailed elsewhere \cite{Pauw-2013a}. Data is corrected for natural background radiation, transmission, sample thickness, measurement time, primary beam flux, parasitic background, polarisation, detector solid angle coverage and sample self-absorption using in-house developed data reduction software. Deadtime correction is unnecessary at the count rates encountered in this study \cite{Kraft-2009}, and dead-pixel and flatfield corrections are performed by the detector acquisition software prior to ingestion by the data reduction software \cite{Eikenberry-2003}. The intensity is subsequently binned using 200 bins linearly spaced over the aforementioned $q$ range, and scaled to absolute units using a calibrated glassy carbon standard \cite{Zhang-2010a}. Statistical uncertainty on the data points is set to the maximum of either: 1) The propagated Poisson counting statistics, 2) The sample standard deviation in the histogram bin, or 3) 1\% of the intensity in the histogram bin as described in \cite{Pauw-2013a}. As the parasitic scattering is minimal due to complete beam path evacuation, improved statistics have been obtained by skewing the time division between sample and background measurement to favour sample measurement time \cite{Pauw-B2012}. Furthermore, the minimal parasitic scattering improves the data accuracy by its reduced sensitivity to inaccuracies in the transmission factor determinations during background subtraction.

Data analysis has been performed using the MC method described in Pauw et al. \cite{Pauw-2013}, replacing the sphere scattering functions with the scattering from radially isotropically oriented rods of fixed aspect ratio. 
This method calculates the scattering from an ensemble of rods, where the scattering from each rod is defined as \cite{Pedersen-1997,Helfer-2005,Pauw-2010a}:
\begin{equation}\label{eq:Prod}
P(q,\phi)=\mid F \mid ^2 = \int\limits^{\pi}_0 \int\limits^{\pi/2}_0  \left[   \frac{2J_1(C)}{C} \mathrm{sinc}(D)    \right]^2 \mathrm{d}\varphi  \mathrm{d}\phi 
\end{equation}
where $F$ is the form factor of the cylinder, $J_1$ is the first order Bessel function, $\phi$ is the cylinder axis orientation projected onto the plane perpendicular to the extrusion direction (i.e. parallel to the detector plane), and $\varphi$ is the out-of-plane degree of tilt of the cylinder axis (c.f. Figure \ref{fig-ebsd-direction}). Furthermore, $C$ and $D$ are defined as:
\begin{equation}\label{eq:rodhelp}
\begin{array}{rl}
C&=qr_2 \sin (\phi) \\
D&=qr_2 \mathrm{AR} \cos(\varphi) \cos(\phi) \\
\end{array}
\end{equation}
with $\mathrm{AR}$ the aspect ratio, defined as the ratio between the rod long axis half-length $r_1$ and rod radius $r_2$: $AR=r_1/r_2$. Note the lack of $\sin\phi$ factor in Equation\,\ref{eq:Prod}, which is reserved for use in spherically isotropic systems but should be omitted for radially isotropic systems. Furthermore, this simple description is valid \emph{only} in cases where the aspect ratio of the scatterers is large, \emph{and} the out-of-plane tilt distribution narrow and centred around the plane normal to the extrusion direction. Given these assumptions, the effect of tilt on the scattering pattern can be approximated as a reduction in projected length of the scatterers. Given the narrow distribution of $\varphi$, the numerical integration is performed over a number of equal-probability segments that make up the probability distribution function. 

The MC method described in \cite{Pauw-2013} calculates a scattering pattern $I_\mathrm{MC}(q)$ using the sum of a fixed number $N_s$ of these elementary scattering contributions (cylinders), through:
\begin{equation}\label{eq:I}
I_\mathrm{MC}(q)=b + A \sum\limits^{N_s}_{k=1} P_k(q) V_k^{2*f_c}
\end{equation}
where $b$ is the constant background level, $V_k$ is the volume for cylinder $k$, and $f_c$ a scaling compensation usually set to 0.5 to suppress the volume-dependency of the relative contribution of $k$ to the ensemble. This is compensated for through the scaling factor $A$, which is defined as:
\begin{equation}\label{eq:A}
A=\nu \Delta\rho^2 \sum^{N_s}_{k=1} \frac {1}{V_k^{f_c}}
\end{equation}
where $\nu$ is the volume fraction of scatterers in the material and $\Delta\rho$ the scattering contrast.

In the starting configuration, the cylinder radii are picked at random within bounds defined by the $q$-limits of the measurement. A goodness-of-fit value is calculated between the scattering pattern $I_\mathrm{MC}$ of this random initial state and the measured scattering pattern, taking the measurement uncertainties and potential flat background level into consideration \cite{Pedersen-1997}. Subsequently, the iterative Monte Carlo procedure is started, which recalculates the goodness-of-fit value after changing the radius of one of the cylindrical contributions\footnote{In the algorithm, this is done by subtracting from the total scattering pattern, the scattering behaviour from one of its constituent contributions. The scattering pattern of a contribution with a different radius is then added.}. 
If the change improves the goodness-of-fit value, the change is kept, otherwise reverted. This process continues until the goodness-of-fit value reaches a value of 1 (or lower), indicating that the set of cylinders at that point effect a scattering behaviour that can describe the measured data \emph{on average} to within the uncertainty of the measured data \cite{Pedersen-1997}. The whole optimisation is restarted several times, and from the variance in these multiple results, uncertainty estimates on the resulting size distributions are determined. Furthermore, a minimum detection limit is determined through the concept of ``observability'' \cite{Pauw-2013}, defined as the minimum required quantity of a scatterer of a given size to make a \emph{measurable impact} on the total scattering pattern $I_\mathrm{MC}$ (i.e. exceeding the measured data uncertainty). 

In other words, scattering in the aged samples is modelled assuming that the only scattering sites are \betap precipitates with [0001] habit, with a fixed aspect ratio and a uniform radial distribution normal to the X-ray beam. 
The aspect ratios (AR) used for the rods are derived from the average aspect ratio measured via TEM observation (See Section~\ref{sec-results-prec-size}). The distribution width of the out-of-plane tilt integration ($\varphi$) is taken from the EBSD results (See Section~\ref{sc:texture}). 
In the ST condition rod-like precipitates are assumed to be absent and scattering functions for spheres are used instead.
The number of contributions $N_s$ has been set to 200 based on the method of maximum efficiency (as described in the MC method paper), and the compensation parameter $f_c$ has likewise been set to 0.5. 
The \betap precipitates have been shown to consist of domains of 
the MgZn$_2$ Laves phase and the Mg$_4$Zn$_7$ monoclinic phase  \cite{RosalieSomekawa2010,SinghPhilMag2010} and the scattering contrast is assumed to be the contrast between Mg and an equal measure of MgZn$_2$ and Mg$_4$Zn$_7$, with densities of 1740, 5160 and 4790\,kg/m$^3$, respectively, resulting in a scattering contrast $\delta\rho^2$ of 6.20$\times10^{30}$\,m$^{-4}$. Thickness of the samples at the irradiated location was calculated using the X-ray transmission factor and overall alloy attenuation coefficient. While the relative uncertainty estimate between datapoints of a measurement as described above can approach as 1\%, the absolute values for the total volume fraction determination is generally considered to be correct to within $\sim$10\% \cite{Pauw-2013a}. 

In places where size distribution peak parameters were determined (i.e. total volume fraction, mean, variance, skew and kurtosis), such determinations were limited to the radius range between 1 and 10\,nm. The lower limit of 1\,nm was selected to coincide with the lower limit dictated by the SAXS measurement range, which is approximately identical to the lower size limit of detection in TEM given the applied magnification. The upper limit of 10\,nm was selected to prevent ``contamination'' of the \betap precipitate radius information with the onset of the rod length information, $\beta_2^\prime$ precipitates and small coarse precipitates whose sizes start appearing beyond this range. In order to fully investigate these three aspects further, ultra-SAXS experiments are required which are beyond the capabilities of the currently available instrumentation. 

\subsection{Transmission electron microscopy}
TEM foils were prepared from post-analysis SAXS samples.
Discs of 3\,mm diameter were punched from the SAXS samples, ground to a thickness of $\sim$50$\,\mu$m and thinned to perforation using a Gatan ion polisher.

Precipitate size and number density were measured using diffraction contrast transmission electron microscopy (TEM) using a JEM 2100 instrument operating at 200\,kV. 
Foils were analysed at ageing times of 6, 24,192 and 406\,h. 
The precipitate length and diameter were measured simultaneously for each precipitate from micrographs recorded with the electron beam aligned normal to the hexagonal axis, (i.e. with the precipitate rods viewed side-on).
The volume of each precipitate was then estimated, assuming each to be a cylindrical rod.
The precipitate half-lengths and radii were binned in the same manner as the SAXS data and weighted by multiplication by the volume of each precipitate. This provided volume-weighted precipitate size data for comparison with the SAXS results.
Images were also recorded with the electron beam along the hexagonal axis (i.e. with the precipitates in cross-section) for the 24\,h and 406\,h samples to confirm that the precipitate diameters measured with the precipitate rods viewed side-on  were consistent.

The TEM observations provide number-weighted distributions of the precipitate size, rather than the volume-weighted distribution derived from SAXS measurements\footnote{SAXS patterns show contributions weighted by their respective surfaces, as detailed elsewhere \cite{Pauw-2013}. Displaying the distributions as volume-weighted distributions is the closest commonly used distribution type amenable to comparisons with other techniques.}.
The TEM precipitate half-lengths and radii were binned in the same manner as the SAXS data and multiplied by the volume of each precipitate.This converts the TEM data to a volume-weighted form for comparison with the SAXS results.

\section{Results}

\subsection{Mechanical properties}
\begin{figure}[htbp]
	\begin{center}
	\includegraphics{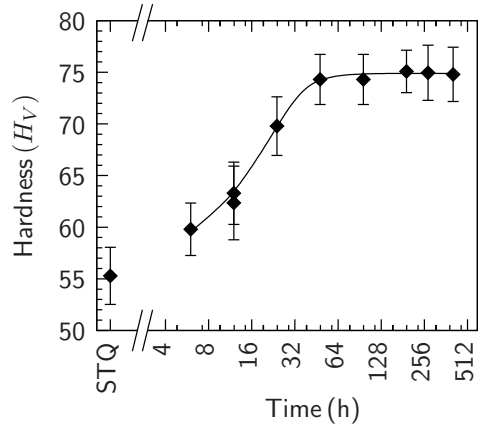}
	\caption{The microhardness of Mg-3.4at\%Zn as a function of ageing time at 150\,$^\circ$C.
Error bars indicate one standard deviation. 
A line has been included as a guide for the eye. \label{fig-hardness}}
	\end{center}
\end{figure}

The alloy has a Vickers microindentation hardness of 55\,$\pm$2 $H_V$ in the solution-treated state (ST).
This increased to 59\,$\pm$2 $H_V$ after 6\,h ageing and underwent a rapid increase Between 12\,h and 48\,h ageing to reach a plateau of 74$\pm$2\,$H_V$. 
The change in hardness with ageing time is shown in Figure~\ref{fig-hardness}.
Error bars indicate the standard deviation in the hardness value, which was in the range of $\pm2\,H_V$.
There was no further significant change in the hardness up to at least 406\,h of ageing.  

\subsection{Microstructure and texture}\label{sc:texture}

\begin{figure}[htbp]
	\begin{center}
	\hfill
	\subfigure[\label{fig-light-micro-TD}]{\includegraphics[width=0.45\textwidth]{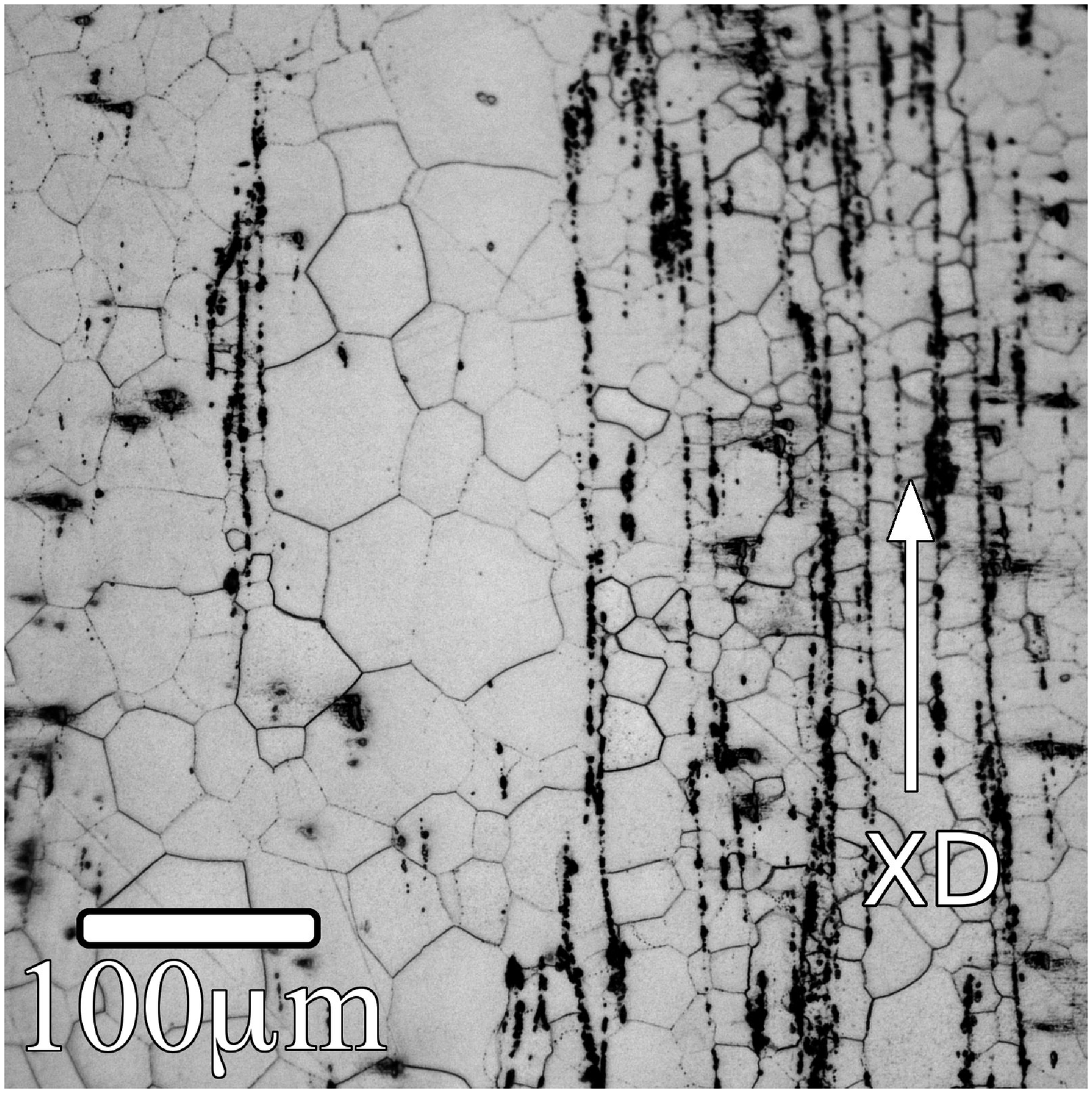}} \hfill\
	\subfigure[\label{fig-ebsd-pf}]{\includegraphics[width=0.45\textwidth]{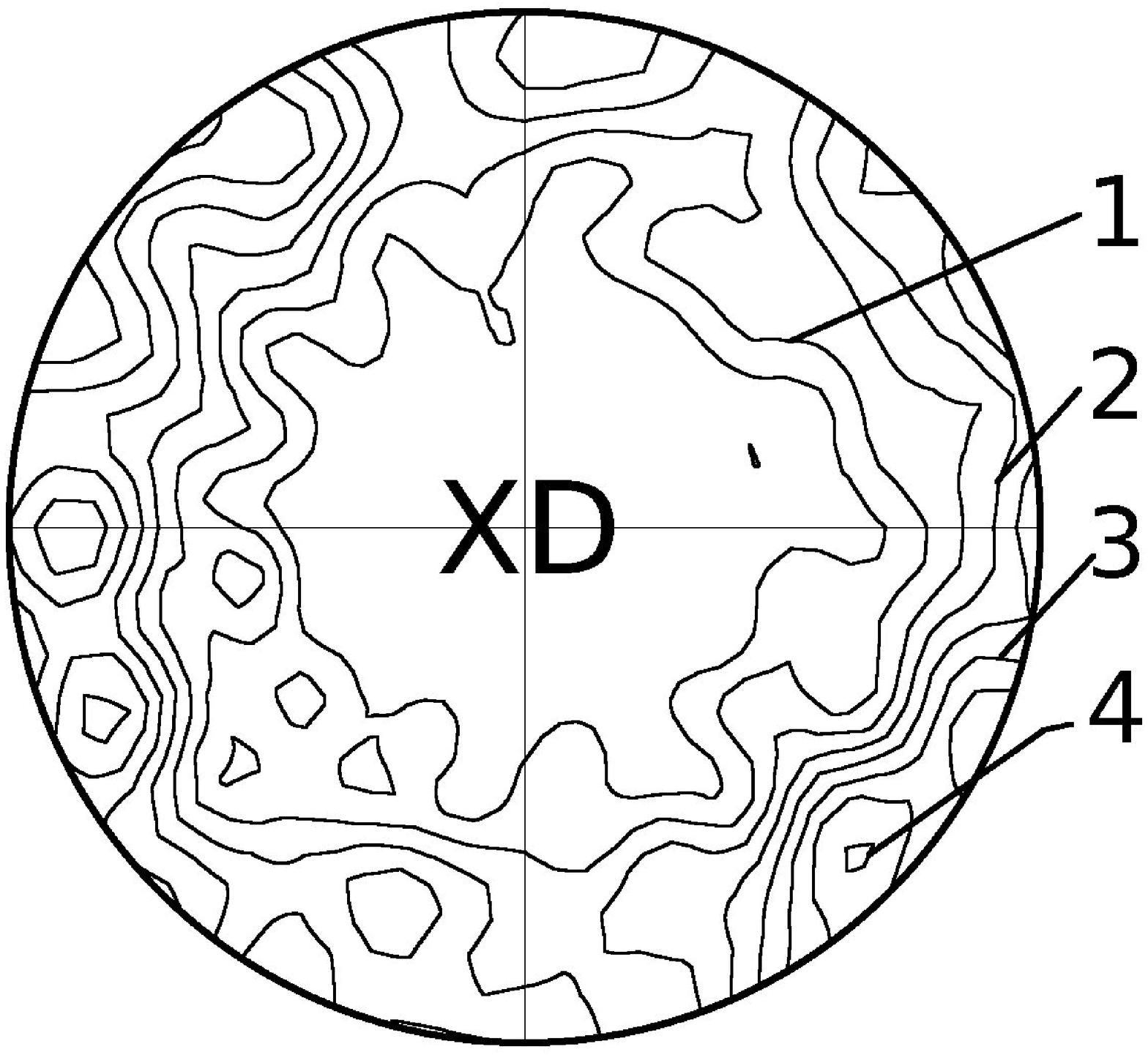}}\hfill
	\subfigure[ \label{fig-sample-geometry}]{\includegraphics[width=0.45\textwidth]{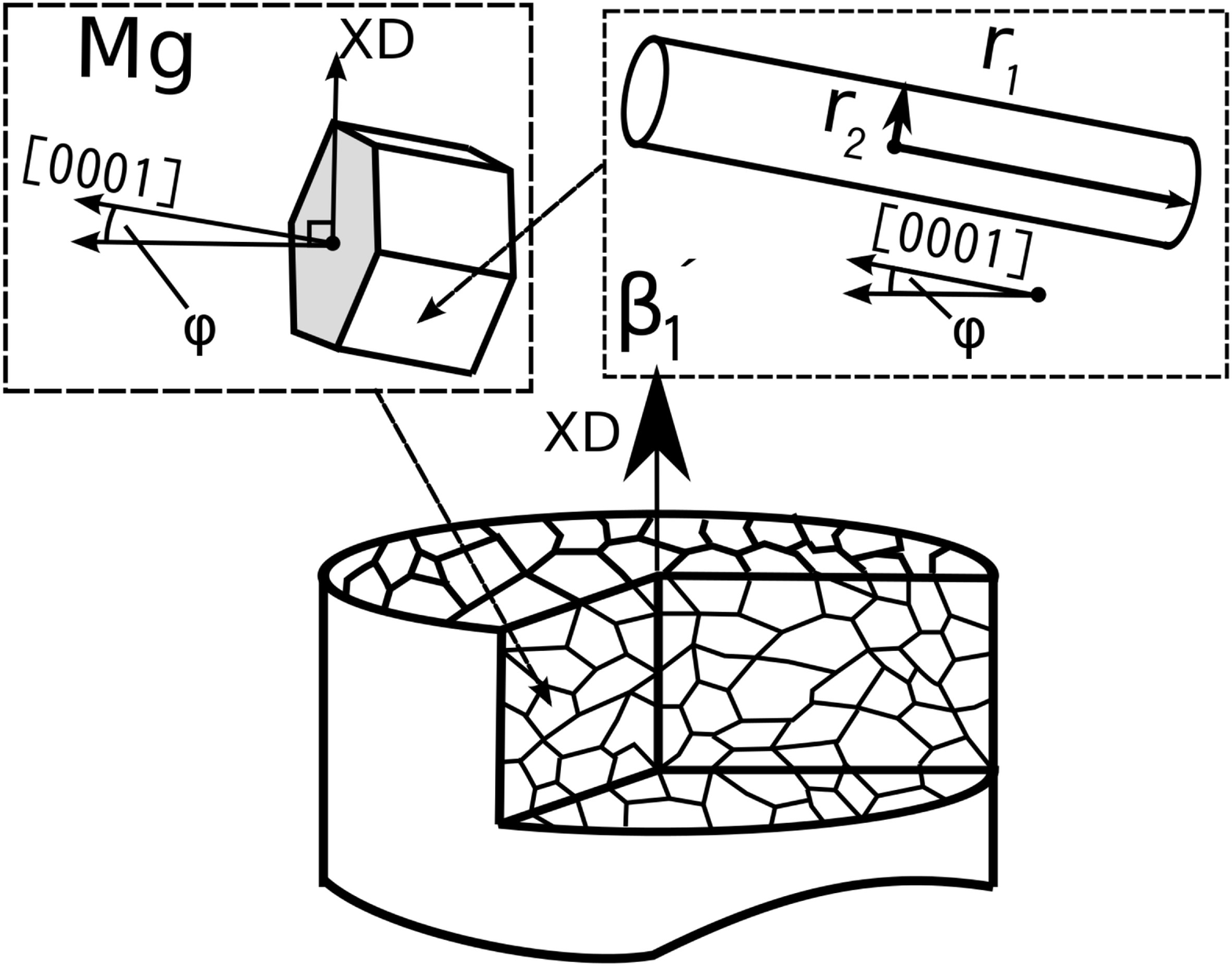}}\hfill\
	\subfigure[\label{fig-ebsd-direction}]{\includegraphics[width=0.45\textwidth]{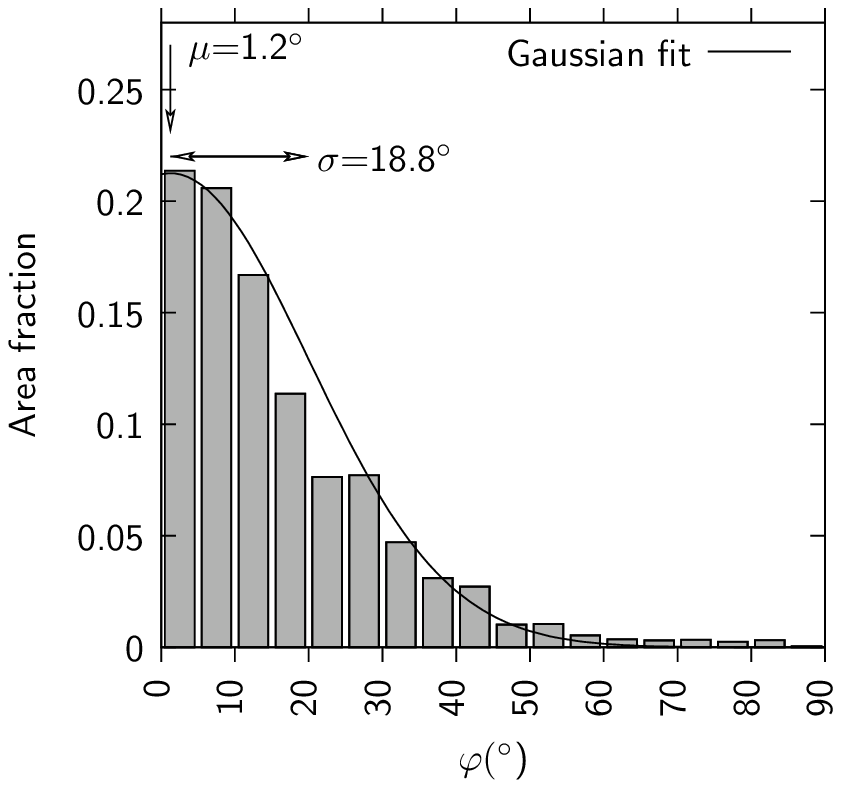}}

	\caption{The texture of the extruded Mg-Zn alloy.
(a) Reflected light micrograph of the alloy, showing the structure normal to the extrusion direction (indicated by arrow). 
(b) Shows a stereographic pole figure obtained from electron back scattered diffraction (EBSD). 
Contours indicate the density of the distribution of [0001]$_\mathrm{Mg}$ poles.
(c) Illustrates schematically the orientation of the \betap precipitates relative to the extruded sample. 
$\varphi$ is defined as the angle between the [0001] axis of a given Mg grain and the plane normal to the XD.
The long axis of the rods is parallel to the  [0001]$_\mathrm{Mg}$ direction, such that $\varphi$ also defines the angle between this axis and the plane normal to the XD.
(d) Shows a frequency plot of the out-of plane angle of the \betap rods. A Gaussian fit gives a mean of 1.2$^\circ$ and a standard deviation of 18.8$^\circ$.} 
	\end{center}
\end{figure}

Reflected light micrographs shows that the alloy contains equiaxed grains with a grain size of approximately 20\,$\mu$m.
Micron-size particles are present, and when viewed in longitudinal section (Figure~\ref{fig-light-micro-TD}) it is clear that these particles are distributed in bands parallel to the extrusion direction.
 
Based on EBSD maps, the average grain size is 17\,$\mu$m. 
The alloy has a rod texture\cite{TurnerTome1994}, as expected for extruded Mg-Zn,
with the  [0001] axes of the Mg grains radially distributed normal to the XD, as shown in the pole figure provided in  Figure~\ref{fig-ebsd-pf}.
The extrusion direction lies at the centre of the stereogram, whereas the strongest intensity for [0001] poles of $\sim$4 is around the edges, i.e. at 90$^\circ$ to the XD. The EBSD intensities are dimensionless values, and are scaled relative to the intensity for a random distribution (defined as 1.0).

The relationship between the orientation of the bulk, extruded material and the \betap rods is shown schematically in Figure~\ref{fig-sample-geometry}. 
The angle between the cross-sectional plane and the [0001] axis of a given grain is termed $\varphi$.
Previous reports indicate that  \betap precipitates form with the long axis parallel to the [0001] axis, so that $\varphi$ also defines the angle between this plane  and the long axis of the rods.
The  distribution of $\varphi$ angles is calculated from the EBSD data using and shown in Figure~\ref{fig-ebsd-direction}.
A Gaussian fit to the EBSD data shows that the [0001] poles of the majority of the Mg grains (and the \betap precipitates therein) lie  close to the cross-sectional plane, with the distribution centred at a  mean angle of  1.2$^\circ$ with a standard deviation of 18.8$^\circ$. This texture analysis is used to define the geometry used for the Monte Carlo SAXS data analysis procedure described in the following sections. 

\subsection{Nanostructure} 

\subsubsection{Precipitate morphology}

\begin{figure}
	\begin{center}
	\hfill
	\subfigure[6\,h \label{fig-tem-examples-6}]{\includegraphics[width=0.4\textwidth]{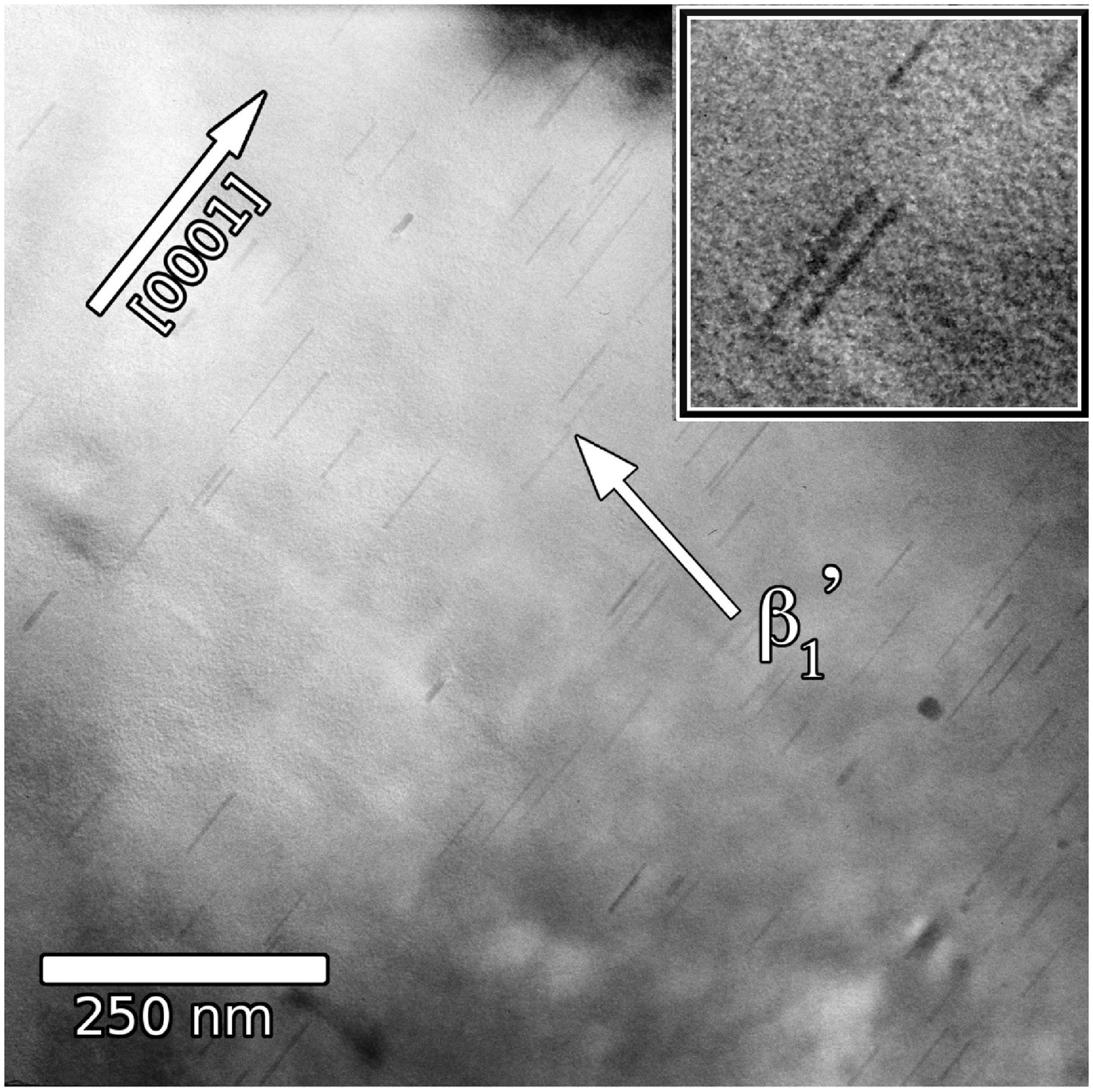}}\hfill
	\subfigure[48\,h.  \label{fig-tem-examples-48}]{\includegraphics[width=0.4\textwidth]{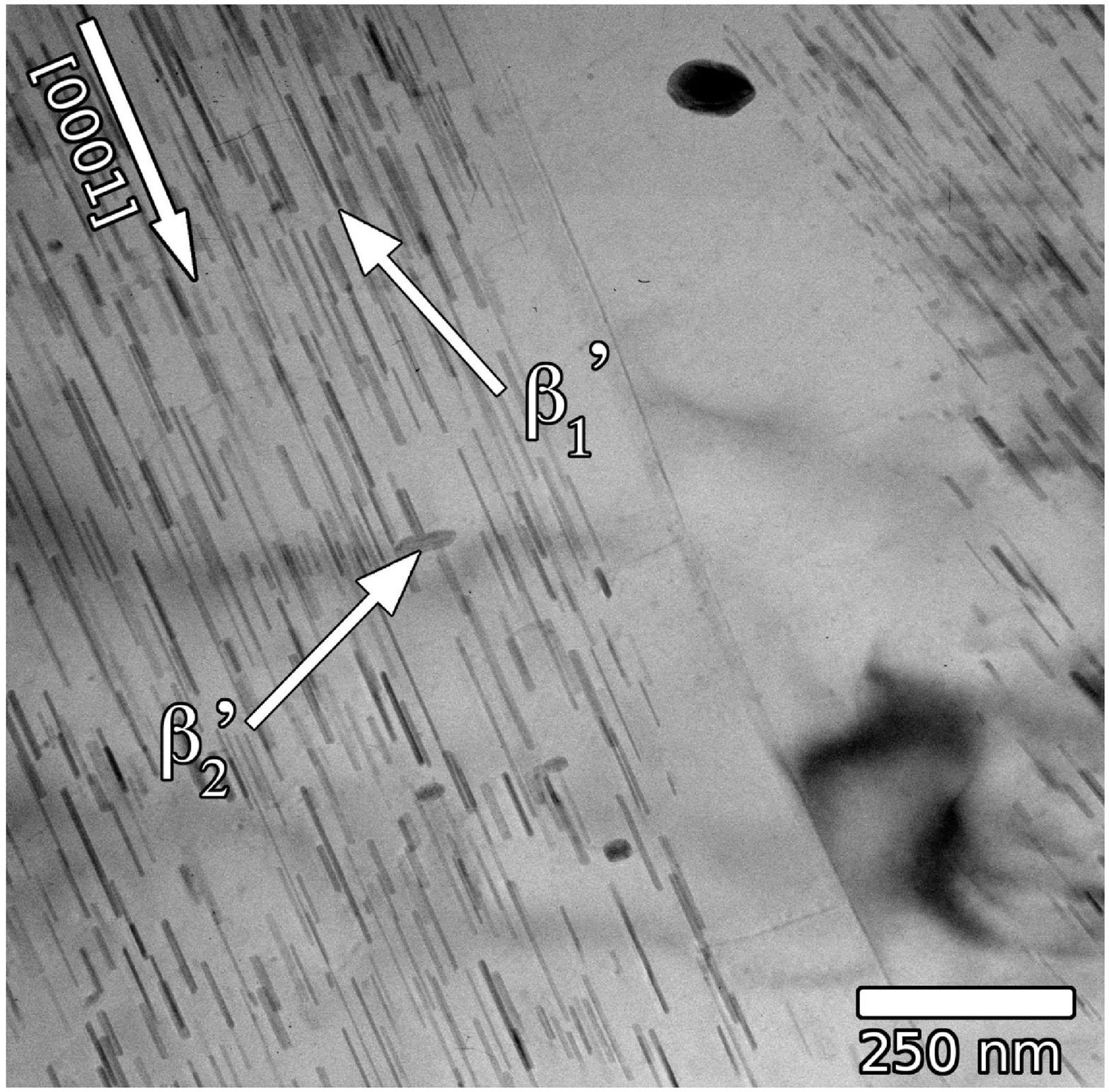}}\hfill\
	\subfigure[406\,h\label{fig-tem-examples-406}]{\includegraphics[width=0.4\textwidth]{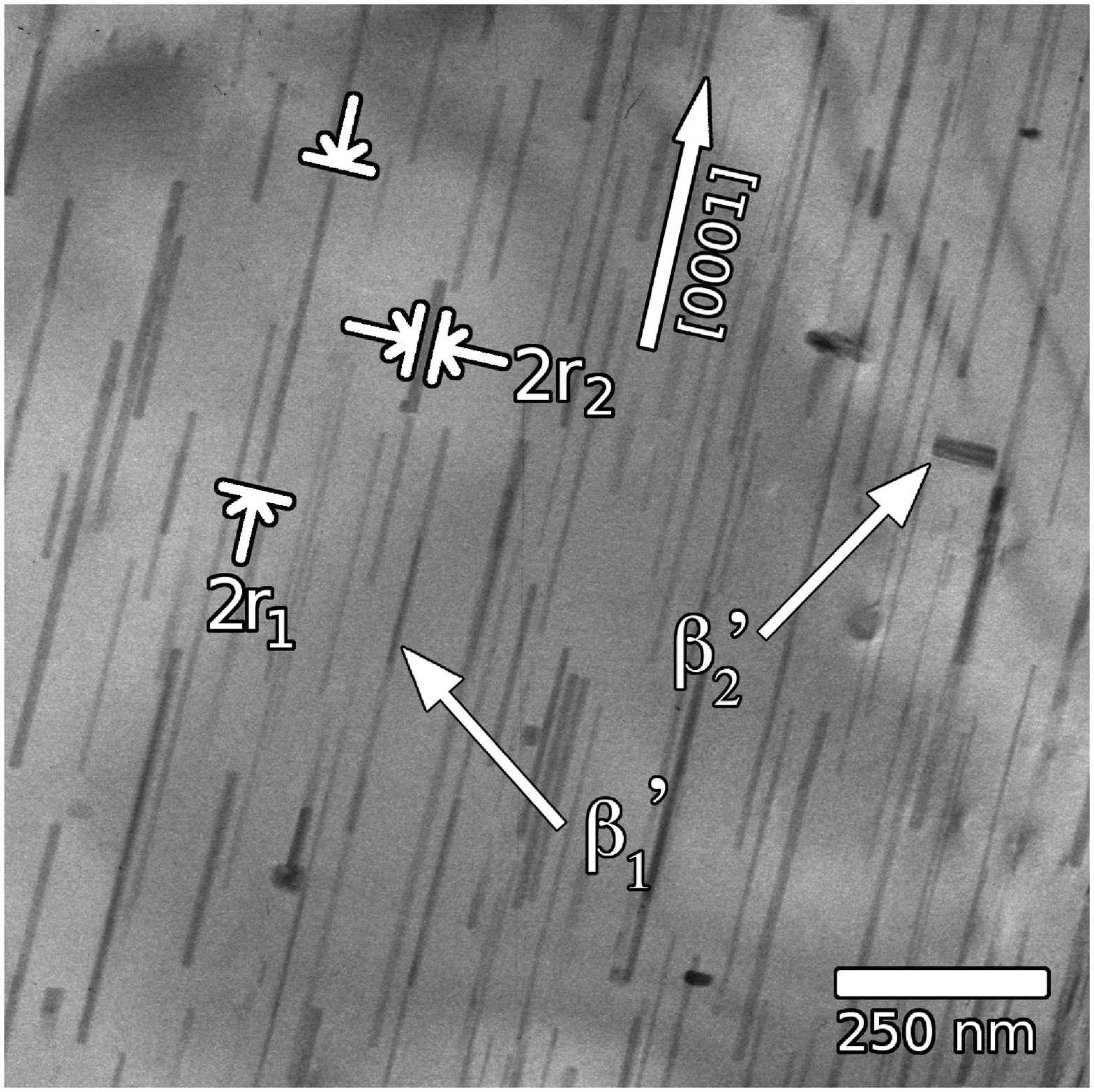}}
	\caption{(a-c) Electron micrographs showing the size and number density of the rod-like $\beta_1^\prime$ precipitates after isothermal ageing time at 150$^\circ$C for (a) 6\,h, (b) 48\,h and (c) 406\,h. 
A small number of plate-shaped $\beta_2^\prime$ precipitates with (0001) habit can also be seen in (b) and (c).
\label{fig-tem-examples}}
	\end{center}
\end{figure}

TEM micrographs show that the principal constituent of the intragranular precipitate nanostructure are rod-like \betap precipitates.
Representative micrographs showing the microstructure at different stages of ageing are shown in Figure~\ref{fig-tem-examples}.
The alloy contains a sparse distribution of rod-like \betap precipitates after 6\,h ageing as shown in Figure~\ref{fig-tem-examples-6}. 
Precipitate length, diameter and number density increased with ageing time, as can be seen in  Figures~\ref{fig-tem-examples-48} (48\,h ageing) and ~\ref{fig-tem-examples-406} (406\,h ageing).
The micrographs in Figure~\ref{fig-tem-examples} were all recorded with the electron beam normal to the [0001] axis, making it possible to measure the precipitate size and aspect ratio for individual particles.  
Figure~\ref{fig-tem-examples-406} also illustrates the geometry used for the measurements of the particle length ($2r_1$) and diameter ($2r_2$), following the notation set out in Figure~\ref{fig-sample-geometry}.

Plate-shaped $\beta_2^\prime$ precipitates with basal plane habit were also observed in samples aged for $\ge$$48$\,h and are  labelled in Figures~\ref{fig-tem-examples}(b,c).
These precipitates are readily distinguishable from the \betap rods by their orientation and stronger contrast, due to their greater thickness in the beam direction, but form only a minor constituent of the microstructure and have been excluded from further analysis.

The micrograph in Figure~\ref{fig-tem-examples-48} shows a grain boundary region with \betap precipitates in both grains. 
There is a grain boundary precipitate-free region in which the \betap precipitates were absent.
The grain boundaries and triple points were often decorated by coarse, low aspect ratio particles frequently of sizes of $\ge1\,\mu$m. 

\subsubsection{Precipitate size measurements \label{sec-results-prec-size}}

Precipitate size distributions are calculated from both TEM and SAXS data.
Precipitate radius, half-length and aspect ratio are measured directly from TEM micrographs obtained with the electron beam normal to the [0001] axis. 

TEM micrographs showed that \betap precipitates undergo substantial lengthening during ageing, with the average half-length, $r_1$, of $33\pm1$\,nm (6\,h ageing)  increasing to 
$160\pm10$\,nm after 406\,h ageing (Estimates of the uncertainty are based on standard error of means measured over multiple regions of the sample).
The average radius also increases from  $3.1\pm0.2$\,nm after 6\,h ageing (Figure~\ref{fig-tem-examples-6}) to $7\pm1$\,nm after 406\,h ageing.

\begin{figure}
	\begin{center}
	\includegraphics[width=0.6\textwidth]{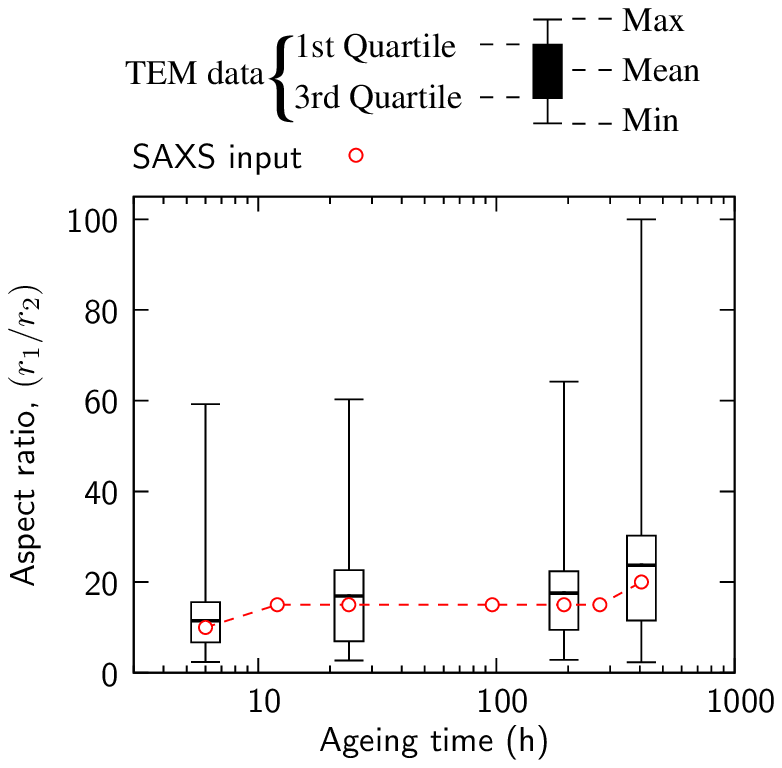}
	\caption{Aspect ratio of the \betap particles as a function of ageing time. Circles indicate the mean aspect ratio, while the error bars show the range of values measured. The open circles indicate the values used as inputs in the SAXS simulation.
\label{fig-tem-aspect}}
	\end{center}
\end{figure}

Figure~\ref{fig-tem-aspect} plots the aspect ratio of the \betap precipitates as a function of ageing time.
The mean aspect ratio increases from 11  after 6\,h ageing to 24 after 406\,h ageing. 
The boxed region of the plot indicates the first and third quartiles of the data, showing that a large proportion of the precipitates have aspect ratios clustered around the mean.
However, there was a broad range between the minimum and maximum values (also indicated on the plot), suggesting that the particles are in various stages of growth. 
Open circles on the plot indicate the aspect ratio values used as input parameters for the SAXS simulation.

\begin{figure}[htbp]
	\begin{center}
	\includegraphics[scale=0.4]{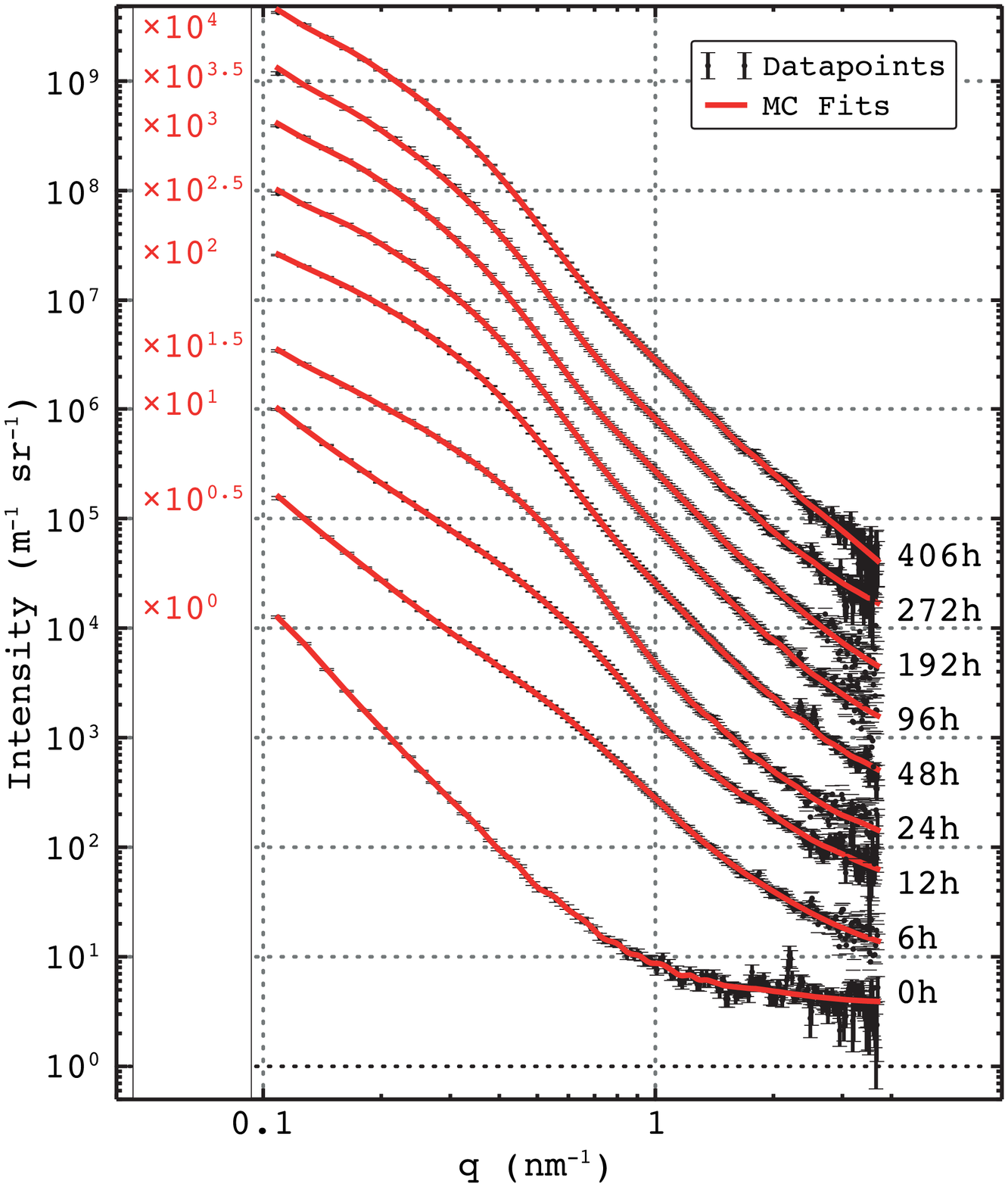}
	\caption{Monte Carlo fits to the mesaured SAXS intensity profiles.
\label{fig-Ifits}}
	\end{center}
\end{figure}

Precipitate size distributions are determined from Monte-Carlo analysis of the SAXS data. 
Only minimal x-ray scattering signal was observed in the solution-treated sample (ST, 0\,h ageing), consistent with the expectation that \betap precipitation requires a finite ageing time and no other scatterers are present in significant proportions. 
In the absence of a known scatterer morphology, by default the scattering was described using spherical precipitates for this sample.  
It should be noted that the coarse precipitates shown in Figure \ref{fig-light-micro-TD} have dimensions beyond the coherence limit of $\sim$200\,nm and thus show no appreciable scattering signal (as discussed in paragraph \ref{sc:saxsmeas}).

\begin{figure}[p]
		\begin{center}
			\includegraphics[height=0.7\textwidth,angle=90]{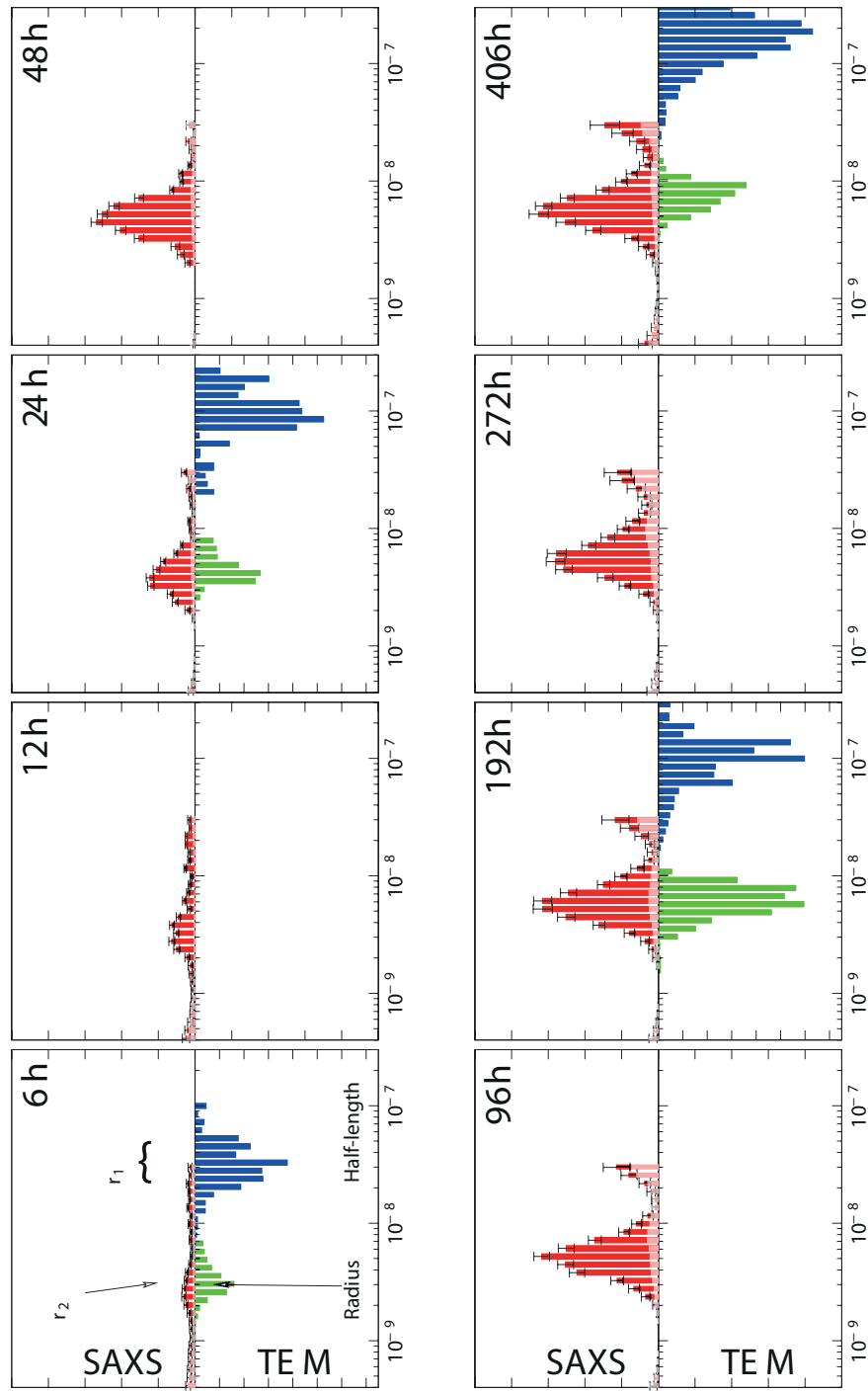}\newline
	        	\caption{Volume-weighted SAXS and TEM particle size distributions for aged Mg-Zn. 
The SAXS distributions were calculated assuming cylindrical rods of aspect ratio 10 (6\,h) or 20 (all other samples). 
The peak labelled $r_2$ corresponds well with the precipitate radii measured in TEM, while the shoulder of a second peak ($r_1$)
appears to coincide with half-length values from TEM observations. The lightened segment of the SAXS distributions indicates the minimum detection limit.
\label{fig-saxs-tem} 
}
		\end{center}
	\end{figure}

The progressive changes in the precipitate size are shown in Figure~\ref{fig-saxs-tem} for SAXS (all ageing times) and TEM (6, 24, 192 and 406\,h ageing).
Both TEM and SAXS data is expressed as volume-weighted particle radii distributions, as this is close to what is actually measured in small-angle scattering measurements \cite{Pauw-2013}. 
For SAXS data the ordinate is the volume fraction, whereas for TEM data it is given in arbitrary units. 
The dominant feature in the SAXS-determined size distribution is a broad peak, initially centred at $\sim 3$\,nm, shifting to an increasing particle size before reaching $\sim 7$\,nm at 406\,h ageing. 
The volume fraction of this component increases rapidly between 6 and 48\,h ageing more gradually for longer ageing times.
The tail of a second peak can also be seen at higher precipitate radii, with the majority of that distribution likely appearing beyond the detection limits of the instrument.
As with the main size distribution component, this tail increases in volume fraction with ageing time.

TEM data for precipitate radius ($r_2$) and half-length ($r_1$) are displayed below the respective SAXS size distribution plots in Figure~\ref{fig-saxs-tem}, showing a remarkable agreement of the distribution of the precipitate radii in TEM and the $r_2$ peak in SAXS.
For ageing times of up to 192\,h the precipitate half-lengths seem to agree with the position and magnitude of the secondary SAXS peak, (labelled $r_1$)  suggesting that this peak corresponds closely with scattering from the long axis of the rods. The limits in the detectable size range of this SAXS instrument ($r_{max} \approx3.5\times10^{-8}$\,m), however, do not allow for determination of the position of the peak centre.

\begin{figure}[htbp]
	\begin{center}
	\includegraphics[width=0.48\textwidth]{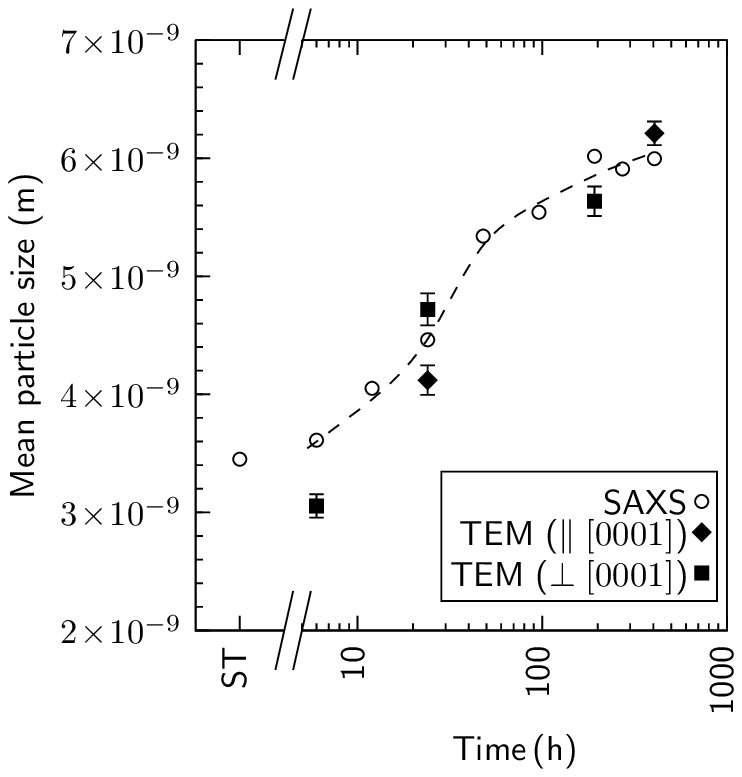}
	\caption{\betap precipitate radius as a function of ageing time. 
The plot shows the mean radius of \betap precipitates as measured by SAXS and TEM.
Error bars for TEM data represent the standard error of means.
TEM measurements made parallel to the [0001] axis (i..e with \betap rods in cross section) gave lower values for the precipitate radius than those made perpendicular to this axis (i.e. with the \betap rods side-on).
 \label{fig-size-evolution}}
	\end{center}
\end{figure}

The average particle radius is plotted versus the ageing time in Figure~\ref{fig-size-evolution}. 
TEM data is converted to volume-weighted values and is shown for measurements  with the rods viewed both side-on and in cross-section. 
The values for SAXS were calculated from  the volume-weighted average for particles in the size range 1-10\,nm, encompassing  the main peak and surrounding region in Figure~\ref{fig-saxs-tem}.
This prevents the average value for SAXS being skewed by additional scattereyrs, including the 
(proportionally more intense) scattering from the long axis of the precipitates.

\subsubsection{Precipitate volume fraction and number density}
\begin{figure}[htbp]
	\begin{center}
	\subfigure[\label{fig-volume-fraction}]{\includegraphics[width=7.5cm]{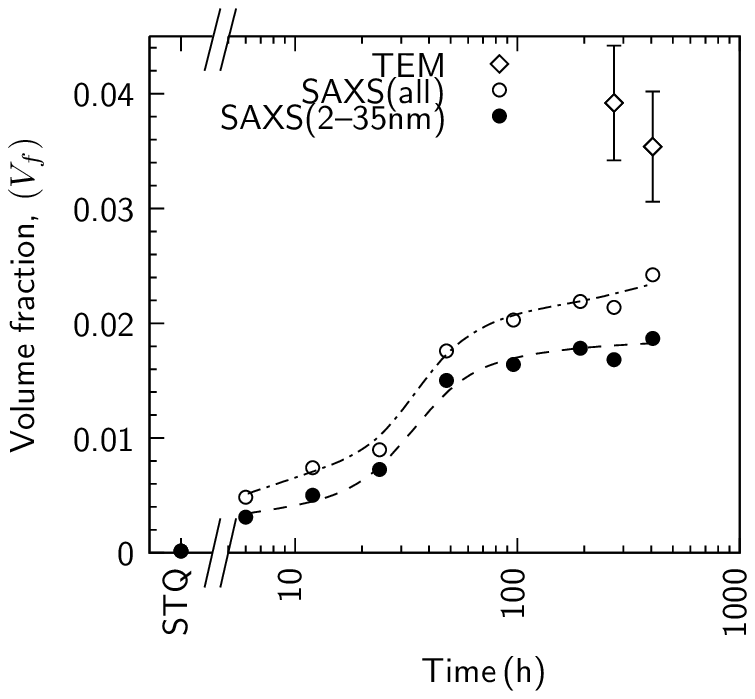}}\hfill
	\subfigure[\label{fig-number-density}]{\includegraphics[width=7.5cm]{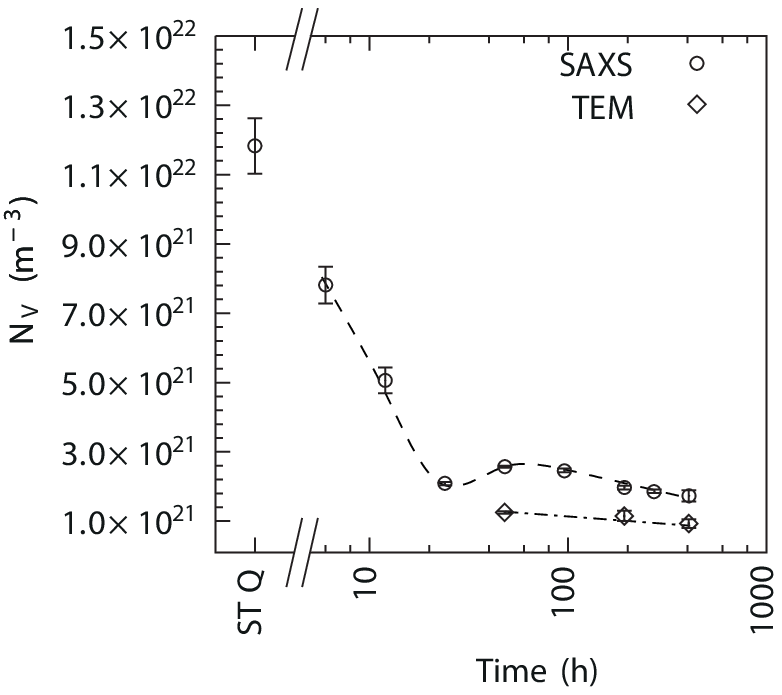}}
	\caption{The (a) volume fraction and (b) number density of \betap precipitates  as a function isothermal ageing time.}
	\end{center}
\end{figure}

The volume fraction of particles estimated from the SAXS data show good qualitative agreement with the age-hardening behaviour. 
Figure~\ref{fig-volume-fraction} shows the volume fractions calculated for the full particle size range amenable to SAXS analysis (2--35\,nm) and also for particles with radii of 1--10\,nm.
The total volume fraction measured by SAXS is approximately 50--100\% greater than the fraction in the 1--10\,nm size range,  possibly due to contributions from the long axis of the rods.
In the ST condition the scatterers are assumed to be spheroidal whereas for the aged samples it was assumed that all particles 
were rods with a given aspect ratio provided by TEM (See Figure~\ref{fig-tem-aspect}).

The volume fraction in the 1-10\,nm range is extremely small in the ST condition ($1.09(4)\times10^{-4}$), reflecting the expectation that \betap precipitates require a finite  ageing time to nucleate and grow. 
A rapid increase in volume fraction occurs between 24\,h and 48\,h ageing $1.43(7)\times10^{-2}$  after which it remains essentially constant. 

Assuming the precipitates are composed of equal amounts of  MgZn$_2$ and Mg$_4$Zn$_7$, the total volume fraction based on the alloy composition and solution treatment temperature is expected to be $\sim$0.032. 
TEM volume fractions calculations give values close to this (e.g. 0.035(5) for 406\,h ageing) in samples where it was possible to tilt the grain to a [0001] orientation.
Precipitate volume fractions measured by SAXS are, however, systematically lower than expected from the alloy composition and TEM measurements. 

The number density of \betap precipitates is calculated from the volume fraction and mean (number-weighted) precipitate volume and shows similar systematic deviations from values measured via TEM. 
In ST condition the  number density (assuming spheroidal scatterers) is estimated at $1.18(7)\times10^{22}$m$^{-3}$. 
The number density decreases to  reach a minimum of $2.09(5)\times10^{21}$m$^{-3}$ after 24\,h, before increasing to  $2.57(3)\times10^{22}$m$^{-3}$ after 48\,h ageing. Further ageing leads to a gradual decrease in particle number density.
Number densities are calculated for TEM  samples where it is possible to tilt the grain to a [0001] orientation and are approximately 50-60\% of the values calculated by SAXS, probably due to the discrepancy in the volume fractions.

\section{Discussion\label{sec-discussion}}

It has been possible to demonstrate remarkably good agreement between the \betap particle sizes (Figure~\ref{fig-size-evolution}) and even the size distributions (Figure~\ref{fig-saxs-tem}) measured via TEM and SAXS, achieving the main goal of this study. This result is particularly significant given that oriented particles with  a wide range of radii and aspect ratios are present. However, there is a consistent discrepancy in the  volume fractions determined (and the  number densities derived from them), however, which needs further consideration.

The reasons for the systematic differences in the volume fraction and number density are not yet clear. The values from SAXS are reliant on the correct determination of a sizeable number of parameters, such as the sample transmission factor and thickness, as well as other calibration factors. Likewise, the volume fraction determined from TEM is  dependent on accurate thickness determination. 
There will be some uncertainty in the scattering factors used  since the proportions of MgZn$_2$ and Mg$_4$Zn$_7$ phases in the precipitates  have not been quantified. 
However, the phases differ in density by approximately 8\% and this would not account the discrepancy in the volume fraction and number density. 
The calculations are also contingent on the precipitate geometry and aspect ratio used, and an increase in the aspect ratio leads to a proportional increase in volume fraction (and corresponding decrease in number density).
SAXS appears to be more sensitive than TEM to particles of radius 1-3\,nm and it is possible that this can lead to higher 
number densities of such particles. 

\subsection{Growth behaviour of \betap rods in Mg-Zn alloys \label{sec-growth}}

The results indicate that \betap precipitation involves two distinct stages, one occurring   between 0--6\,h ageing and the other  between 24--48\,h ageing. 
The first stage involves a twenty-fold increase in the volume fraction of \betap precipitates from $1.09(4)\times10^{-4}$ to $2.41(4)\times10^{-3}$.
This coincides with an increase in the microindentation hardness from 55\,$H_V$ to 59\,$H_V$ and relates to the initial precipitation of \betap particles.

A second distinct stage occurs between 24\,h and 48\,h ageing, during which time the alloy reaches the peak hardness value of $74\pm2\,H_V$.
The number density increases by approximately 25\% during this interval, which, combined with a rapid increase in the particle size, sees a rise in the volume of \betap precipitates to reach a plateau.

The decline in precipitate number density  between 6 and 24\,h ageing indicates that the increase in hardness over this interval can be ascribed  to growth of the existing precipitates rather than further nucleation. This suggests that \betap precipitation is nucleation-limited at 150$^\circ$C, consistent with studies that have shown that  the ageing process in Mg-Zn \cite{RosalieMgZn2012} and Mg-Zn-Y\cite{RosalieMgZnY2013} alloys can be accelerated by introducing microstructural  defects through pre-ageing deformation.

In addition to the rod-shaped, \betap precipitates,  $\beta_2^\prime$ also precipitate during ageing.
These precipitates adopt plate or lath shaped morphologies, with the largest dimension normal to the hexagonal axis.
The appearance of this phase in place of the \betap particles is been associated with the loss of mechanical strength during extended heat-treatment (i.e. overageing).
Scattering from $\beta_2^\prime$  particles would be registered as that of extremely low length, high radius \betap particles (i.e. aspect ratio $\ll1$).
If present in substantial volume fractions this would artificially decrease the average precipitate length and increase the average radius.
However, TEM confirms that $\beta_2^\prime$ precipitates observed at ageing times $\ge48$\,h  remain a very minor constituent even after 406\,h ageing at 150$^\circ$C, justifying their exclusion from the SAXS Monte Carlo model.
The low number density of  $\beta_2^\prime$ precipitates (roughly one to two orders of magnitude lower than the \betap precipitates) makes quantification of their size and volume fraction infeasible. 
However, the observation that the microindentation hardness reached a plateau from 64--406\,h ageing is consistent with this phase having little influence on the microstructure.

\begin{figure}
	\begin{center}
		\includegraphics{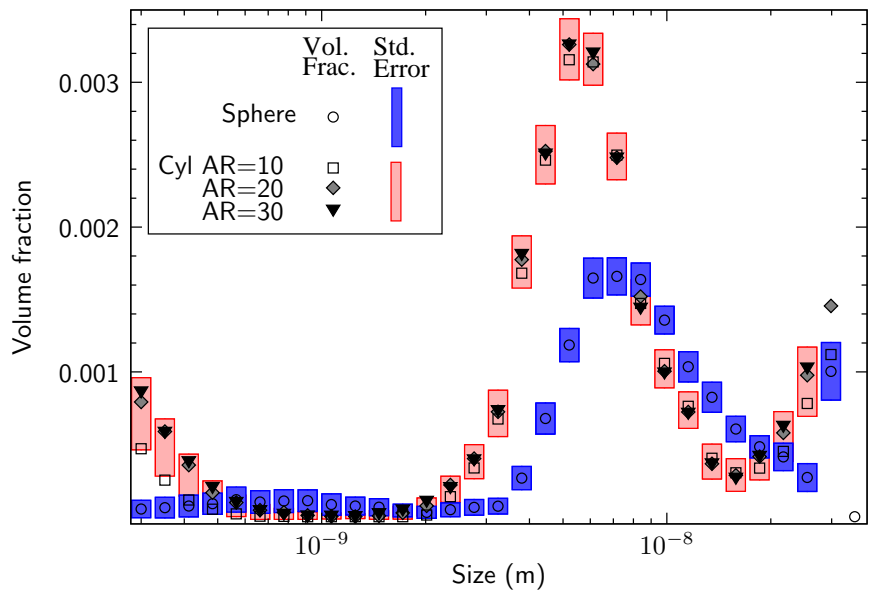}
		\caption{A comparison of SAXS size distributions calculated with different precipitate geometries.
The figure shows the size distributions calculated in the sample aged for 406\,h, modelled using either spheres or cylinders (of aspect ratio of 10, 20 and 30). The scattering from each simulation matched the experimental scattering within the defined convergence limit. 
Open circles show the volume fractions for the spherical model. The open boxes show the standard error for this model.
The volume fractions for cylinders of AR=10, 20 and 30 are indicated by squares, diamonds and triangles,respectively. 
The filled boxes are centred on the mean volume fractions for the three cylindrical models and their size indicates the standard error between  the three aspect ratios.
The calculated size distribution  differs considerably between spheres and cylinders.
However, except at the extreme lower end of the plot, the values for the three cylindrical fits lie within experimental error of one another. 
 \label{fig-aspect-comparison}}
	\end{center}
\end{figure}

In order to highlight the need for TEM information in SAXS analysis, and to indicate that information on the aspect ratio or even precipitate shape cannot be uniquely identified from standalone SAXS, SAXS analyses using deviant geometries have been performed. The Monte Carlo simulations were repeated assuming \emph{spheroidal} scatterers, as well as with rod-like scatterers of aspect ratio 10, 20 and 30, to evaluate the effect of changing the modelled precipitate geometry. These solutions also described the data to within the data uncertainty, reaffirming the need for externally supplied morphological parameters (e.g. as can be provided by TEM) to generate a unique solution. The size distributions obtained from matching the scattering from samples aged for 406\,h with spheres and cylinders of aspect ratio 10, 20 and 30 are presented in Figure~\ref{fig-aspect-comparison}. 
In general, spherical models give a broader, more skewed size distribution than a cylindrical model. The spherical model gives a maximum volume fraction of $1.7(1)\times10^{-3}$ centred at a radius of 7.2\,nm. The three cylindrical models are in good agreement with one another and averaging the fits gives a maximum volume fraction $3.2(2)\times10^{-3}$ centred at a radius of 6.1\,nm.
This demonstrates that, for a cylindrical fitting geometry, a wide range of aspect ratios can fit the scattering within experimental error, and that it is therefore not possible to extract the aspect ratio information from these scattering patterns.

Incorrect selection of the modelling geometry may not change the size distribution but can severely affect other calculations such as the volume fraction calculations. When these are based on spheres, they yield volume fractions of approximately half those for the rod-like particles, with a maximum volume fraction of 0.015(3) after 406\,h ageing. 
The aspect ratio is also crucial in this case since increasing the aspect ratio used for a cylindrical fit will increase the volume fraction.
This is particularly important since scattering from the long axis of the rods falls outside the range of the present SAXS instrument and an inaccurate aspect ratio would not be apparent from SAXS measurements alone. 
Errors such as this could lead to greatly under- or overestimated interparticle distances, which are essential for estimation of the precipitate strengthening effect. 

As the precipitate length is beyond the range of most commonly available SAXS instruments, it is not possible to compare these directly with TEM data.

\subsection{Scattering from anisotropic precipitates}

\begin{figure}
	\begin{center}
		\includegraphics[width=0.7\textwidth]{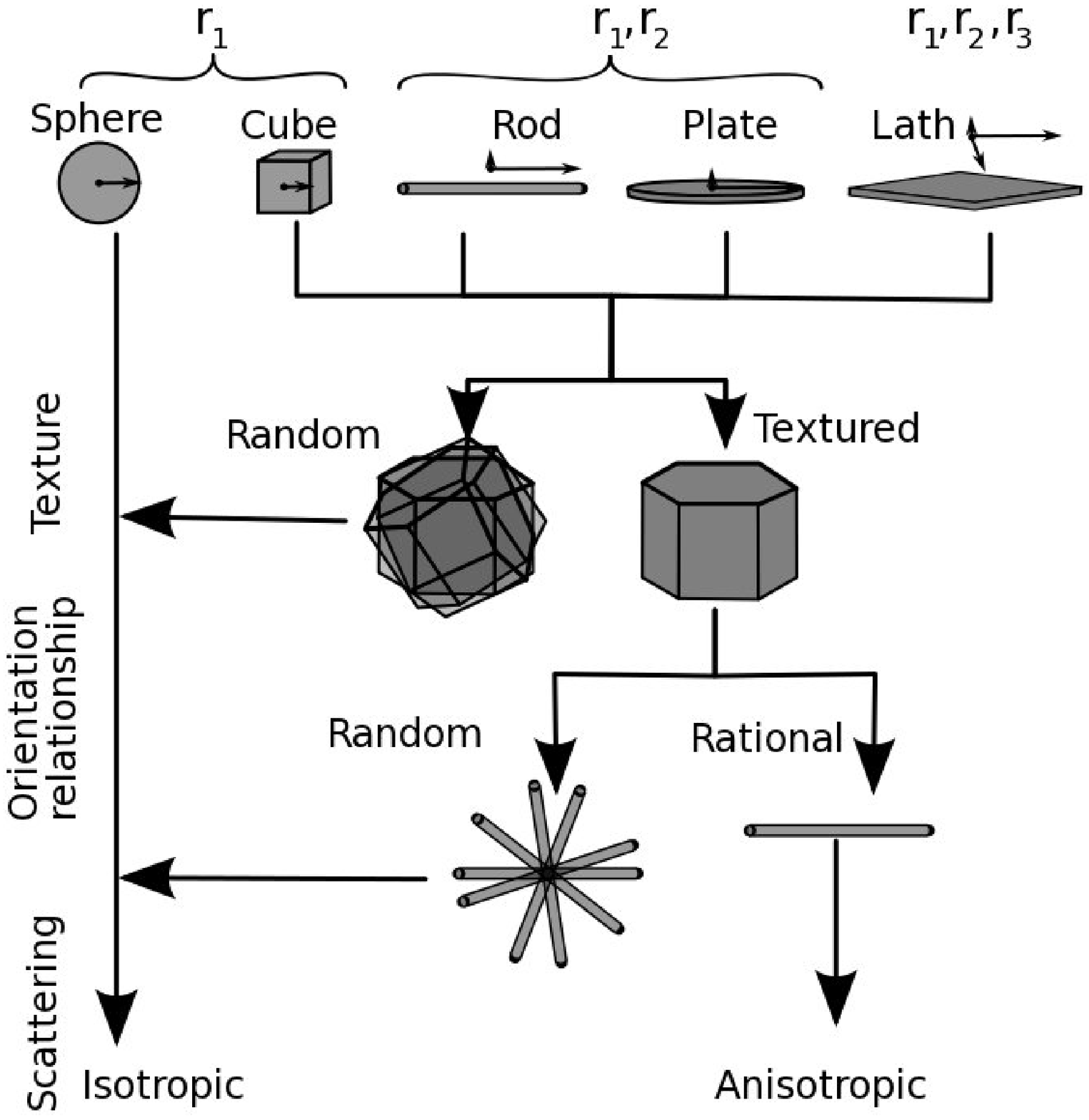}
		\caption{The effect of precipitate orientation relationship and crystallographic texture on the anisotropy of SAXS. 
In order to obtaining 2-d scattering information on precipitates with two or three size parameters ($r_1,r_2\dots$) it must be possible to measure an anisotropic scattering pattern. 
This requires that the material have a non-random crystallographic texture and that the precipitates have a rational orientation to the matrix. \label{fig-anisotropy-chart}}
	\end{center}
\end{figure}

Scattering from anisotropic precipitates in a polycrystalline materials depends not only on the morphology of the precipitate, but on the texture and precipitate habit as well.
Precipitates are generally described by a few basic morphologies (illustrated schematically in Figure~\ref{fig-anisotropy-chart}), which  include spheres and cubes (both described by a single size parameter, radius ($r_1$)).
rods, (described by two parameters: $r_1$ and $r_2$ with $r_1\gg r_2$ and rotational symmetry around $r_1$)
and plates (also radially symmetric about $r_1$ but for which $r_2\gg r_1$). 
Lath-like precipitates are also formed in some systems and require three parameters to describe the size; 
$r_1,r_2,r_3$ with $r_3> r_2>r_1$.

The scattered radiation will be isotropic when the scatterers are spherical, or alternatively when the texture is random and/or the precipitates are randomly oriented with respect to the matrix.
In the latter two cases it is difficult to distinguish the scattering from that of a distribution of spheres, making it the determination of realistic particle sizes problematic, stressing the need for independent determination of the precipitate morphology.
As noted in Section~\ref{sec-growth} it is entirely possible to model the SAXS intensity to the same degree of accuracy assuming spherical precipitates, producing a mathematically correct, but physically unrealistic solution. 

Scattering will be anisotropic for high aspect ratio particles in materials with a non-random texture.
This case is widely applicable since anistropic precipitates generally have well-defined orientation relationship between the precipitate and the matrix and defined crystallographic habit. 
In addition, most alloys show a non-random texture resulting either from oriented growth or from deformation.

Anisotropic scattering makes it possible to extract physically representative particle sizes from high-aspect ratio particles if the relationship between the beam direction and scatters can be clearly defined.

The case of [0001] rods in extruded Mg-Zn is relatively straightforward.
As shown in Figures~\ref{fig-ebsd-pf} and \ref{fig-sample-geometry} the combination of specimen orientation, precipitate-matrix orientation and radially symmetric texture ensure that for the majority of grains the long axis ($r_1$) is perpendicular to the beam. This results in essentially radially isotropic scattering from the  $r_1$ component, greatly simplifying the SAXS analysis.

It is important to note that the texture is a global average over the sample and if the analysis includes either a few larger grains, or is not locally radially isotropic then streaks can arise from the anisotropy in the orientation of the rods. 
If, for example, the x-ray beam was not accurately centred on a radially symmetric sample, it is probable that the scattering would become anistropic. 
One approach to combat this issue would be to translate the sample during the SAXS measurement, effectively further averaging out any local texture. 

\begin{figure}
	\begin{center}
		\includegraphics{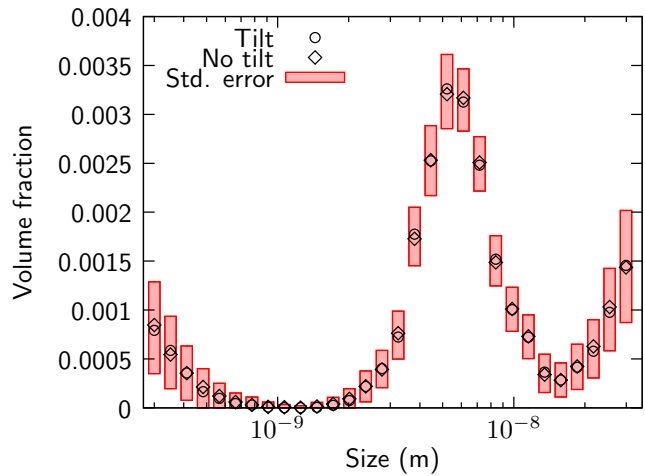}
		\caption{The effect of out-of plane tilting of the \betap precipitates.
Circles and diamonds indicate the size distributions calculated without including tilting of the \betap rods.
The shaded areas represent the standard error of the two methods. \label{fig-tilt-comparison}}
	\end{center}
\end{figure}

Even if such larger-scale variations in texture are accounted for it is important to note that the texture is a statistical average and individual grains may diverge from the ideal orientation.
This was accounted in the present work for by including tilting of the rods in the SAXS simulation, with the orientation distribution based on a Gaussian fit of the EBSD data (Figure~\ref{fig-ebsd-direction}). 
The inclusion of tilting the \betap rods out of the radial plane substantially increases the computational burden of the Monte Carlo fitting procedure and may not be necessary in many cases, particularly in strongly textured materials.
In this alloy, which has a maximum texture intensity of $\sim$4, the inclusion of out-of-plane tilting has little effect on the radii distribution.
Figure~\ref{fig-tilt-comparison} shows the volume-weighted size distributions for Mg-Zn aged for 406\,h analysed using a model with and without the consideration of tilt. 
The shaded areas represent the standard error of the two methods. 
Size distributions calculated with and without the consideration of tilting of the rods are shown by circles and diamonds respectively.
There is very good agreement between the two simulations, which differ by considerably less than the standard error. 
Similar good agreement is found for the size distributions measured at all ageing times.

\subsubsection{Capabilities and limitations of the Monte Carlo SAXS-TEM method} 

The Monte Carlo SAXS method is capable of rapid determination of meaningful three-dimensional information from sample volumes orders of magnitude greater than feasible with TEM. 
TEM particle sizes were based on multiple images of different areas of each foil, with 500 particles measured on average in the [0001] axis direction and 200-300 normal to this direction. 
For a beam diameter of 0.2\,mm and foil thickness of 0.6\,mm the X-ray beam covers a volume of approximately 0.02\,mm$^3$. Based on the number density estimates this encompasses approximately $9\times10^{13}$ precipitates. This number can be greatly increased by simply translating the sample during measurement to average over a larger volume of material.
Furthermore, SAXS can also be used to follow precipitation \textit{in situ}. 

The method presented here is readily applicable to cases where a simple geometric relationship exists between the precipitates and the X-ray beam.
Application of this method is more challenging if the precipitates adopt  more than one orientation relationship and/or have multiple possible morphologies.
While not intractable such complications would make extracting data difficult and necessarily reliant on a detailed understanding of the texture.

Precipitate volume fraction measurements via SAXS require that the phases be accurately identified, since the scattering cross-section is determined by the electron density of the scatterers.
Difficulties may occur due to the presence of additional phases with different compositions, or if a single phase exists over a broad composition range. 
In this instance, \betap particles have been widely reported and well characterised in Mg-Zn based alloys  \cite{RosalieSomekawa2010,Gao2007,Singh2007} and although the composition could feasibly range between Mg$_4$Zn$_7$ and MgZn$_2$ the overall effect on the scattering is minor. 
In the case of unknown or poorly-characterised phases the precipitates present should be identified from bulk x-ray diffraction or other techniques.
In contrast, TEM measurements are determined from the area fraction in projection, but suffer from greater statistical uncertainty due to the inherently microscopic analysis volumes.

A final limitation on the present analysis concerns the particle size range of the SAXS instrument, which can only collect  x-rays scattered from particles up to  35\,nm in radius. Overcoming this restriction would greatly enhance the potential for Monte Carlo analysis of SAXS from  high aspect ratio particles. 
With this objective in mind. a purpose-built Ultra-SAXS apparatus is currently under construction at our institute which would extend the range of particle sizes up to about $2\mu$m. This instrument will greatly extend the utility of SAXS particle size measurements, making it possible not only to measure large isotropic particles but also to simultaneously measure two size parameters in high aspect ratio precipitates.
\section{Conclusion}
The nanoscale precipitation behaviour of a Mg-Zn alloy has been examined in detail, made easy through complementary TEM and SAXS. The method presented here obtains quantitative size distributions and internally consistent volume fractions and number densities of non-globular, high-aspect ratio precipitates and takes into account both the precipitate orientation and texture.  

Accurate analysis by SAXS was contingent upon relating the orientation of the bulk sample to that of the principal scattering sites, which was achieved through texture measurement of an extruded sample via EBSD and subsequent determination of the precipitate morphology and habit using TEM. 
This provided the information necessary for SAXS analysis through Monte Carlo fitting, yielding a quantified precipitate size distribution and volume fraction for the anisotropic rod-like precipitates.

Samples aged  at 150$^\circ$C contained rod-like \betap precipitates, with the size, number density and volume fraction of particles increasing with ageing time. 
The size distributions determined through TEM and SAXS analysis were in very good agreement for ageing times between 6 and 406\,h. 
The precipitate volume fraction underwent an order of magnitude increase during initial ageing (0-6\,h ageing) with a further 16\%  increase between 24\,h and 48\,h ageing. The later increase was coupled with an increase in the particle number density of approximately 25\%, indicating that further \betap nucleation occurred in this interval.

Extended ageing did not result in a deterioration of the hardness and the alloy retained values within experimental error of the peak value up to 406\,h.
This is consistent with the findings that ageing at this temperature does not result in formation of significant amounts of plate-shaped $\beta_2^\prime$ precipitates.



\end{document}